\documentclass[aps,amsfonts,prl,twocolumn,showpacs]{revtex4-1}

\usepackage{pdfpages} 
\makeatletter
\AtBeginDocument{\let\LS@rot\@undefined}
\makeatother

\usepackage{subfigure}
\usepackage{textcomp}
\usepackage{graphicx}
\usepackage{bm}
\usepackage{amsmath,amssymb,amsthm}
\usepackage{amsfonts}
\usepackage{dsfont}
\usepackage[colorlinks=true,linkcolor=blue,urlcolor=blue,citecolor=blue]{hyperref}
\usepackage{multirow}
\usepackage{url}

\begin{document}

\title{The Fredkin staircase: An integrable system with a finite-frequency Drude peak}

\author{Hansveer Singh$^{1}$, Romain Vasseur$^{1}$ and Sarang Gopalakrishnan$^{2}$}
\affiliation{$^1$Department of Physics, University of Massachusetts, Amherst, Massachusetts 01003, USA \\
$^2$Department of Physics, Pennsylvania State University, University Park PA 16802, USA}

\begin{abstract}

We introduce and explore an interacting integrable cellular automaton, the Fredkin staircase, that lies outside the existing classification of such automata, and has a structure that seems to lie beyond that of any existing Bethe-solvable model. 
The Fredkin staircase has two families of ballistically propagating quasiparticles, each with infinitely many species. 
Despite the presence of ballistic quasiparticles, charge transport is diffusive in the d.c.~limit, albeit with a highly non-gaussian dynamic structure factor. 
Remarkably, this model exhibits persistent temporal oscillations of the current, leading to a delta-function singularity (Drude peak) in the a.c.~conductivity \emph{at nonzero frequency}. 
We analytically construct an extensive set of operators that anticommute with the time-evolution operator; the existence of these operators both demonstrates the integrability of the model and allows us to lower-bound the weight of this finite-frequency singularity.


%

\end{abstract}

\maketitle

\emph{Introduction}---
Conventional hydrodynamics predicts that high-temperature transport in lattice systems should generically be diffusive. 
This expectation is strongly violated in two tractable classes of systems. First, kinetically constrained models (KCMs)---i.e., models of stochastic dynamics subject to hard kinetic constraints, initially introduced as toy models of glassy behavior~\cite{GARRAHAN2018130,Fredrickson:1984tl,https://doi.org/10.48550/arxiv.1009.6113,Ritort_2003,Everest:2016wv}---have been shown to exhibit subdiffusive transport under quite general conditions. In some subdiffusive KCMs the constraints lead to the presence of additional conservation laws, giving rise to ``fracton hydrodynamics''~\cite{Iaconis:2019vd,Feldmeier:2020tq,Gromov:2020vu,Morningstar:2020tw}; in other cases, it is unclear whether additional conservation laws are present~\cite{Singh:2021vy,Richter:2022tg}. Subdiffusion in KCMs was experimentally observed in Ref.~\cite{PhysRevX.10.011042}. A second class of systems that escape conventional hydrodynamics are integrable systems~\cite{PhysRevX.6.041065,PhysRevLett.117.207201,PhysRevLett.119.220604,Bastianello_2022}, which have extensively many conservation laws, as well as stable ballistically propagating quasiparticles. One might expect transport in integrable systems to be ballistic, but in the presence of internal symmetries, integrable spin chains can exhibit transport that is sub-ballistic, and either diffusive or superdiffusive~\cite{PhysRevLett.121.160603,Gopalakrishnan:2018wd,PhysRevLett.122.127202,10.21468/SciPostPhys.6.4.049,Bulchandani_2021}. Remarkably, there are deep links between integrable systems and KCMs: if one applies the update rules of a KCM in certain deterministic sequences (rather than at random) one obtains discrete-time integrable cellular automata. This correspondence has been noted in multiple cases,  see {\it e.g.}~\cite{Medenjak:2017ty,Gopalakrishnan:2018wj,Gopalakrishnan_2018,PhysRevLett.123.170603,Bu_a_2021,Pozsgay_2021,Gombor:2021ww,10.21468/SciPostPhys.12.3.102}; how general it is, and how the properties of the stochastic and integrable versions of the model are related, remain open questions.

\begin{figure}[!t]
	\includegraphics[width=\columnwidth]{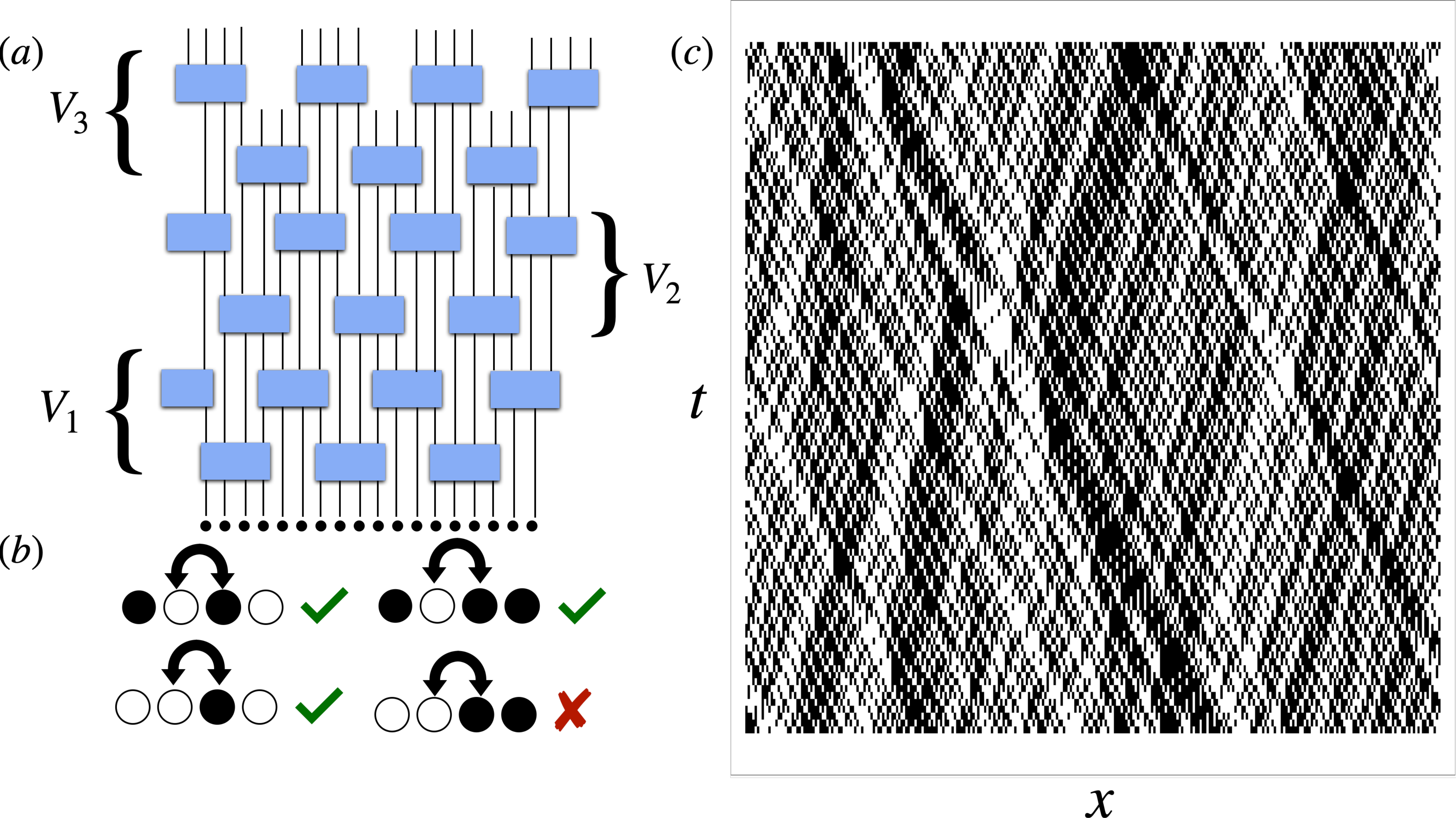}
	\caption{{\bf Model.} (a) Pictorial representation of the circuit geometry described by Eq.\ref{eqn:layer}. (b) Pictorial representation of the rules associated to the Fredkin constraint as described in Eq.\ref{eqn:gates}. $\bullet$ represents particles while $\circ$ represents holes. (c) Time evolution of a random product state in the occupation basis.}
	\label{fig:geomrules}
\end{figure}

In the present work we identify a new such correspondence, by constructing an integrable cellular automaton based on the Fredkin update rule. This update rule acts on four adjacent sites, swapping the middle two provided the outer sites obey the kinetic constraint sketched in Fig.~\ref{fig:geomrules}. The stochastic Fredkin model~\cite{Singh:2021vy,causer2022slow} has nontrivial subdiffusive dynamics, with a spacetime scaling $x \sim t^{3/8}$ whose origin is not yet understood. We construct a cellular automaton with the Fredkin update rule; we call this model the Fredkin staircase automaton (FSA). We show that the FSA is integrable by explicitly showing how an extensive family of conservation laws can be constructed; however, we have not completely solved the model as the conservation laws we construct do not exhaust its quasiparticle content. Because the FSA is a cellular automaton, we are able to extract both the spectrum of quasiparticles and scattering phase shifts between them from numerics. We find two families of quasiparticle species, called $\beta$ and $\sigma$ quasiparticles. The $\beta$ quasiparticles also form bound states, which we call $\beta$-strings; there is an infinite family of these. The scattering properties between $\beta$ and $\sigma$ quasiparticles are unusual, and it is not clear how to relate the integrable structures we find to standard Bethe ansatz concepts. 

In addition to identifying the quasiparticle structure, we study transport in this model by numerically computing its a.c.~conductivity through the Kubo formula~\cite{Bertini:2021tr}. 
Our most unexpected finding is that the a.c.~conductivity has a $\delta$-function ``Drude'' peak at a \emph{nonzero} frequency, associated with persistent oscillations of current fluctuations. We are unaware of any other integrable systems with a nonzero-frequency Drude peak (although related oscillations have been observed in non-hydrodynamic quantities~\cite{PhysRevB.102.041117}). 
On the other hand, the d.c.~limit of the conductivity is finite, so transport is asymptotically diffusive despite the presence of ballistic quasiparticles. 
In addition, the dynamical structure factor of this model is spatially strongly asymmetric, and obeys a scaling form $C(x,t) = t^{-1/z} f(x/t^{1/z})$, with $z=2$ and $f$ a skewed, non-Gaussian scaling function. We discuss the origin of these transport phenomena in terms of an infinite family of charges that {\em anti-commute} with the time evolution operator.

\emph{Model}.--- Our system is a one dimensional chain of qubits of length $L$ whose basis states we represent as $|\bullet\rangle$ and $|\circ\rangle$ to denote whether a particle has occupied a site or not. The dynamics is governed by a Floquet operator $\mathcal{U}$, shown pictorially in Fig.~\ref{fig:geomrules}, which is composed of three layers of four site unitary gates, i.e. $\mathcal{U}=V_{3}V_{2}V_{1}$, where
\begin{equation}
	V_{i} = \prod_{j\equiv i\,\text{mod}\,3}U_{j,j+1,j+2,j+3},
\label{eqn:layer}
\end{equation}and
\begin{equation}
\begin{split}
	U_{j,j+1,j+2,j+3} &= P^{\bullet}_{j}\text{SWAP}_{j+1,j+2}P^{\bullet}\\
	&+P^{\bullet}_{j}\text{SWAP}_{j+1,j+2}P^{\circ}_{j+3}\\
	&+P^{\circ}_{j}\text{SWAP}_{j+1,j+2}P^{\circ}_{j+3}\\
	&+P^{\circ}_{j}\mathds{1}_{j+1,j+2}P^{\bullet}_{j+3}.
\label{eqn:gates}
\end{split}
\end{equation}
$P^{\bullet}_{j} = |\bullet\rangle\langle\bullet|_{j}$, $P^{\circ}_{j} = |\circ\rangle\langle\circ|_{j}$ and $\text{SWAP}_{j,j+1}$ is the usual swap gate. Note that these gates locally preserve particle number so that the total particle number of the system is conserved.
 We remark that the gate geometry is equivalent to a staircase geometry hence the name: Fredkin Staircase Automaton, as the constrained swaps satisfy the so-called Fredkin constraint~\cite{PhysRevB.94.155140,Chen_2017,Chen:2017vl,PhysRevB.104.115149,Udagawa_2017,Zhang_2017,https://doi.org/10.48550/arxiv.1605.03842,Chen:2020wb,Movassagh_2016,Salberger_2017,Langlett:2021uk}. The gate geometry leads to an asymmetric circuit light-cone for local operators where the support of the light cone increase by 6 sites to the left and 3 sites to the right. Owing to the fact that the circuit light cone increases by a multiple of 3, we will partition our chain into unit cells containing 3 sites. We note that the gate pattern we are using is crucial for the model to be integrable. In the supplemental material \cite{SM}, we show that deforming the gate geometry breaks the integrability of the model and leads to subdiffusion with an exponent $z\simeq 8/3$ in line with the predictions of Ref.~\cite{Singh:2021vy}. Each update conserves the total number of $\bullet$ (and $\circ$) sites, so we can regard the fraction of $\bullet$ sites as the ``filling fraction'' $f$.\par

\emph{Quasiparticles}.--- We first identify single quasiparticle excitations of the FSA model above its vacuum state (i.e., the state $|\circ\rangle^{\otimes N}$). One can create states with a single elementary quasiparticle by occupying a single site. Since there are three inequivalent sites in the unit cell there are three inequivalent quasiparticles. For the gate pattern and unit cell in Fig.~\ref{fig:geomrules}, quasiparticles on the first and third sites of the unit cell move ballistically leftward with velocity $v_{\sigma}=3/2$, whereas those on the second site move rightward with velocity $v_{\beta} = 3$---as this notation anticipates we will call the two left-moving quasiparticles $\sigma$-particles and the right-moving quasiparticle a $\beta$-quasiparticle. (We will avoid calling them left- and right-movers as the direction they move is set by the arbitrarily chosen chirality of the gate pattern.) 

We now turn to states with two occupied sites. When the occupied sites are more than one unit cell apart, such states have two quasiparticles, with flavors that depend on the location of the filled site in the unit cell (as above). When the occupied sites are in the same or adjacent unit cells, the identification is more subtle. One might have expected that a configuration like $\bullet\circ\bullet\circ\circ\circ$ or $\circ\circ\bullet\bullet\circ\circ$ contains two $\sigma$-quasiparticles; in fact they contain a $\sigma$ and a $\beta$ quasiparticle in the process of colliding. In this sense, $\sigma$-quasiparticles obey a hardcore constraint: if one slot is occupied its neighbor cannot be. Moreover, as we will see below, two $\beta$-quasiparticles in adjacent unit cells are not independently propagating, but instead form a bound state. 

\begin{figure}[!t]
	\includegraphics[width=\columnwidth]{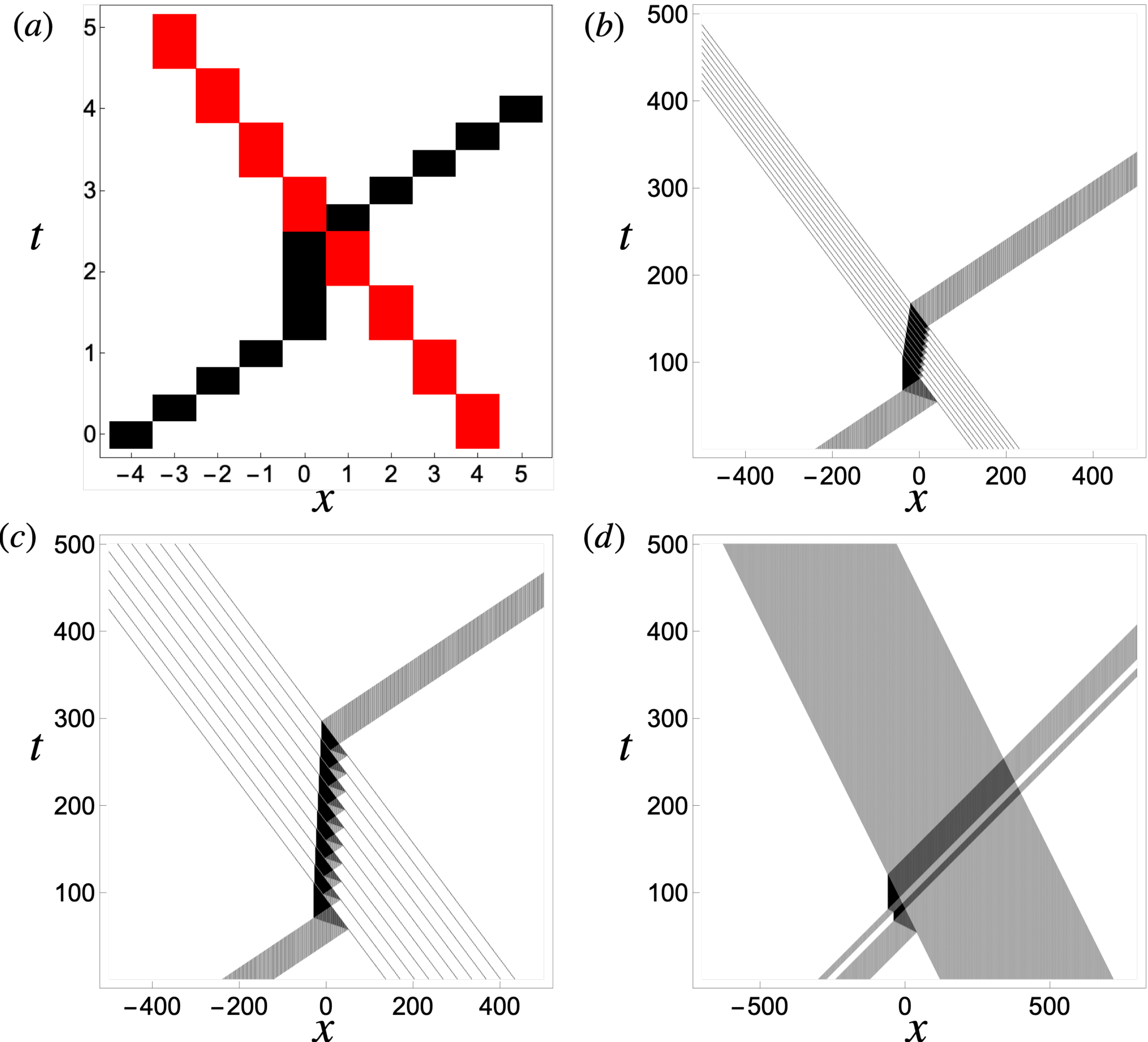}
	\caption{{\bf Quasiparticles scattering.} (a) A collision between a $\sigma$ (colored red) and $\beta$ (colored black)  particle where integer time steps represent evolving by a full Floquet step while fractional steps indicate evolving by individual layers. One can see the $\sigma$ particle receives no scattering shift but the $\beta$ particle is delayed by one Floquet time step. (b) A size 40 $\beta$-string (moving right) collides with a size 200 2-spaced type-A $\sigma$-string (moving left). One can clearly see that the velocity of the $\beta$-string is renormalized when it is inside the $\sigma$-string. (c) A size 40 $\beta$-string collides with a size 200 10-spaced type-A $\sigma$-string. Observe that the the $\beta$-string's velocity is much lower than compared to the previous situation indicating that the effective velocity of $\beta$-strings is highly dependent on spacings of $\sigma$-particles . (d) Two $\beta$ strings collide when they encounter the large $\sigma$-string. One can see that the smaller $\beta$-string overtakes the larger one after the collision occurs.}
	\label{fig:strings}
\end{figure}

We now turn to the scattering between quasiparticles. Here, in contrast to standard integrable systems, we find a strong asymmetry between $\sigma$ and $\beta$ quasiparticles: the trajectories of $\sigma$ quasiparticles are totally unaffected by collisions, while $\beta$ quasiparticles are slowed down. When colliding with a single $\sigma$ quasiparticle, a sequence of $s$ $\beta$ quasiparticles is slowed down by $s$ unit cells. These sequences thus form collectively moving bound states, which we call $\beta$-strings of size $s$; collisions with $\sigma$ quasiparticles renormalize the velocities of such $\beta$-strings in an $s$-dependent way. 
Lastly, we note that because $\beta$-strings of different sizes have different renormalized velocities (in the presence of $\sigma$ strings), two $\beta$-strings can collide (when a smaller string tries to overtake a larger one). 
In the bottom-right panel of Fig.~\ref{fig:strings}, we show a collision between a size 40 $\beta$-string and size 10 $\beta$-string. The two collide with each other once they encounter the $\sigma$-string and then one observes that the smaller $\beta$-string ``overtakes" the larger $\beta$-string. When two $\beta$-strings of size $(s,s')$ collide the faster of them is further sped up (and the slower is further slowed down) by $2\min(s,s')$ unit cells. This scattering phase shift precisely parallels the result for Heisenberg spin chains.\par

To set up generalized hydrodynamics for this model, we would need the scattering shifts between an arbitrary-size $\beta$-string and an arbitrary configuration of $\sigma$ quasiparticles. In a typical Bethe-ansatz solvable problem, one would have to identify all the distinct $\sigma$-type strings and the scattering shift accumulated by a $\beta$-string passing through the $\sigma$ quasiparticles would be a sum of shifts due to each $\sigma$-type string. In the FSA this separation does not happen: rather, the scattering shift is sensitive to the separation between $\sigma$ quasiparticles (Fig.~\ref{fig:strings}). Consider, as a simple example, the case of a $\beta$-string of size $\ell$ scattering off two $\sigma$ quasiparticles separated by $d$ unit cells: the resulting scattering shift is $\min(2 \ell, \ell + 2d)$ unit cells. Thus, from the point of view of their scattering properties, even two arbitrarily well separated $\sigma$ quasiparticles cannot be treated as independent scatterers with additive scattering shifts. Although we are able to find expressions for the scattering shift of an arbitrary $\beta$-string in an arbitrary background of $\sigma$ quasiparticles, it is not clear how to express these in the standard Bethe ansatz form. Nevertheless, our numerical results strongly suggest that all quasiparticles are stable (so the model is integrable), and we now explicitly demonstrate this for the $\sigma$ quasiparticles.

\begin{figure}[!t]
	\includegraphics[width=\columnwidth]{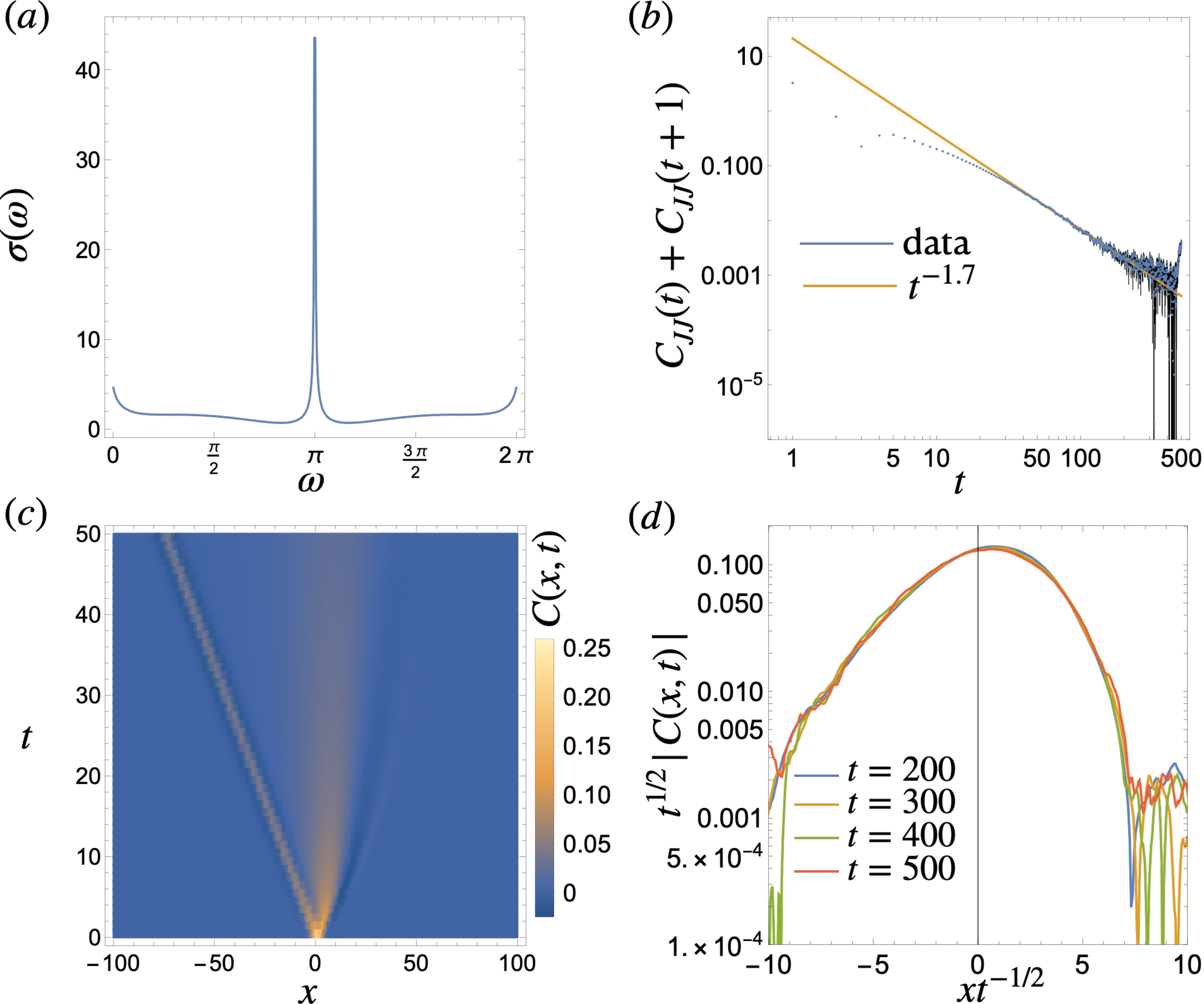}
	\caption{{\bf Transport.} (a) {a.c.}~conductivity $\sigma(\omega)$. Note that $\omega=\pi$ features a prominent peak indicating a $\pi$ Drude weight and that we also have a finite non-zero value at $\omega = 0$ which suggests diffusive behavior. (b) Behavior of the Kubo correlator $C_{JJ}(t)+C_{JJ}(t+1)$ which is twice the average value over one period of the oscillations in the current-current correlator. $C_{JJ}(t)+C_{JJ}(t+1)$ falls off in a power law fashion, i.e. $t^{-\beta}$, with $\beta \approx 1.7 >1$ indicating the presence of a finite non-zero diffusion constant at low frequency. (c) Behavior of the particle structure factor $C(x,t)$ at short times and (d) Diffusive scaling collapse of the structure factor, $C(x,t) = t^{-1/2} f(x/t^{1/2})$ with $f$ a non-Gaussian skewed scaling function. 
	(a) and (b) is data averaged over $10^{8}$ realizations and (c) and (d) are averaged over $10^{7}$ realizations.} 
	\label{fig:current}
\end{figure}
\emph{Integrability}.--- In this section we show that there are an infinite number of quasi-local operators that are conserved. The construction is based on the intuition that operators which correspond to creating $\sigma$-strings should be rather simple to construct since $\sigma$-strings propagate without scattering delays. First, we note that a $\sigma$-particle is given by the one of the following configurations: $\bullet\square\circ$ or $\square\square\bullet\circ\square\square$, where $\square$ is meant to denote that a particle may or may not occupy that site. 
Using the above fact, one can construct a number operator counting the total number of $\sigma$-particles spaced by $s$ unit cells given by
\begin{equation}
	N_{s} = N_{s}^{\bullet\times\circ}+N_{s}^{\bullet\circ},
\end{equation}
where
\begin{align} \label{Eqcharges}
N_{s}^{\bullet\times\circ}&=\sum_{i\equiv1\text{mod}3}P^{\bullet\times \circ}_{i}\prod_{j=i+1}^{i+s}(\mathds{1}-P_{j}^{\bullet\times \circ})P^{\bullet\times \circ}_{i+s+1},  \\
	N_{s}^{\bullet\circ}&=\sum_{i\equiv1\text{mod}3}P^{\bullet}_{i-1}P^{\circ}_{i}\prod_{j=i+1}^{i+s}(\mathds{1}-P_{j-1}^{\bullet}P_{j}^{\circ})P_{i+s}^{\bullet}P_{i+s+1}^{\circ} \notag,
\end{align}
is conserved. If $s\leq 0$ one should replace the product of operators with an identity operator and we note that for $s=-1$, $N_{s}$ simply counts the total number of $\sigma$-particles. We further note that the $N_{s}^{\bullet\times\circ}$ and $N_{s}^{\bullet\circ}$ correspond to the asymptotic spacings \cite{Gopalakrishnan:2018wd} of type-A and type-B $\sigma$-particles, respectively. \par 
One can show that $N_{s}$ is conserved by noting that the Floquet evolution operator, $\mathcal{U}$, maps $P_{i}^{\bullet\times\circ}\mapsto P_{i-1}^{\bullet}P_{i}^{\circ}$
	and 
	$P_{i-1}^{\bullet}P_{i}^{\circ}\mapsto P_{i-3}^{\bullet\times\circ}$ which implies $N_{s}^{\bullet\times \circ}\leftrightarrow N_{s}^{\bullet\circ}$ at each time step. \par 
	All operators commute with each other since they are diagonal in the occupation basis. Additionally, they are orthogonal to each other under the Hilbert-Schmidt inner product (i.e., $\langle A, B \rangle \equiv 2^{-L} \langle A^{\dagger}B \rangle$) because for $s'>s$, all terms in $N_{s'}$ have larger support than all terms in $N_{s}$. Since we constructed an infinite set of linearly independent conserved quasi-local operators, the FSA model is integrable. We note that there are clearly more operators which are conserved such as the total number of $\beta$-strings. It would be interesting to further investigate the algebraic integrable structure of this model in future work~\cite{Gombor:2021ww,Pozsgay_2021,2021arXiv210601292P,10.21468/SciPostPhys.12.3.102,Bu_a_2021,2022arXiv220502038G}.  
	
	\par 
	
\emph{Transport}.--- Because of its integrability, it is natural to expect particle transport in the FSA model to be ballistic: if the particle current overlaps with any of the conserved charges, it cannot fully relax leading to persistent currents. In what follows, we argue analytically and numerically that transport in the FSA is a lot more exotic and interesting: none of the charges uncovered above overlap with the current operator, and we find numerically that low frequency transport is {\em diffusive}. However, we identify analytically another set of changes which {\em anticommute} with the time evolution operator, and which do have a finite overlap with the current. We argue that this leads to a finite Drude peak in the conductivity at frequency $\omega = \pi$. Alternatively, it shows that the FSA is a (fine-tuned) equilibrium discrete {\em time-crystal}~\cite{Khemani:2016vp,PhysRevLett.117.090402, 2019arXiv191010745K,PhysRevB.102.041117}, as it exhibits persistent oscillating currents. 

We characterize the transport properties of the FSA by computing the current-current correlation function, $C_{JJ}(t) = \frac{1}{L}\langle J(t) J(0) \rangle$, where $J(t)=\sum_{x}j(x,t)$ and $j(x,t)$ represents the local current density and $\langle A \rangle \equiv 2^{-L} \text{tr}(A)$ for an operator $A$.  We present the details of the calculation of $j(x,t)$ and its lengthy expression in the supplemental material \cite{SM}. We numerically computed $C_{JJ}(t)$ using classical evolution and averages are performed over $10^{8}$ random initial states. \par 

From the current-current correlator, we compute the~a.c.~conductivity, $\sigma(\omega)$ by using the Kubo	 formula~\cite{Bertini:2021tr}
\begin{equation}
\sigma(\omega) = \frac{1}{2}C_{JJ}(t=0) + \sum_{t=1}^{\infty} {\rm e}^{i \omega t} C_{JJ}(t).
\label{eqn:freqcond}	
\end{equation}
Because of the Floquet (discrete time) nature of model, we have $\omega \in [0, 2 \pi)$. We computed this conductivity numerically, see Fig.~\ref{fig:current}. One can see a clear peak at $\omega = \pi$ indicating persistent oscillations in the time-dependent conductivity and hence also the current-current correlator. We attribute these persistent oscillations to the existence of an extensive number of operators $Q$ such that $\mathcal{U}^{\dagger}Q \mathcal{U} = -Q$.  
To see that such operators imply such persistent oscillations , consider the $\pi$-Drude weight, defined as 
	\begin{equation}
		\mathcal{D}_{\pi} = \lim_{t\rightarrow\infty}\frac{1}{t} \sum_{\tau =1}^t (-1)^{\tau}C_{JJ}(\tau).
	\end{equation}
	The $\pi$-Drude weight characterizes the weight of a possible Drude (delta function) peak in the conductivity at frequency $\pi$.
	
One can show that if a collection of operators $Q_{s}$ satisfy the aforementioned conditions then one can lower bound $\mathcal{D}_{\pi}$ through the application of a Mazur bound \cite{MAZUR1969533,DHAR2021110618,https://doi.org/10.48550/arxiv.2112.12747}, i.e.
	 \begin{equation}
	 	\mathcal{D}_{\pi} \geq \frac{3}{L}\sum_{s} \frac{\langle J(0) Q_{s} \rangle^{2}}{\langle Q_{s}^{2}\rangle}.
	 	\label{eqn:oscMazur}
	 \end{equation}
A family of such operators $Q_{s}$ are given by $Q_{s} = N_{s}^{\bullet\times\circ}-N_{s}^{\bullet\circ}$ and they anticommute with $\mathcal{U}$ since $N_{s}^{\bullet\times \circ}\leftrightarrow N_{s}^{\bullet\circ}$ at each time step:
	\begin{equation}
\lbrace Q_{s} , \mathcal{U}\rbrace = 0
	\end{equation}

 We remark that if these were all the charges which anticommuted with $\mathcal{U}$ then Eq.~\ref{eqn:oscMazur} would become an equality. The fact that $\langle J(0) Q_{s}\rangle \neq 0$ means that $\mathcal{D}_{\pi}$ is non-zero which implies that $C_{JJ}(t)$ necessarily has to be of the form $C_{JJ}(t) = (-1)^{t} (\mathcal{D}_{\pi} + \text{sub-leading terms})$. 
Such persistent oscillations indicate that the FSA is a discrete time crystal---albeit fine-tuned rather than generic~\cite{Khemani:2016vp,PhysRevLett.117.090402, 2019arXiv191010745K,PhysRevLett.128.100601,PhysRevB.102.041117}.   \par 

Despite this exotic behavior near $\omega=\pi$ frequency, low-frequency transport appears to be diffusive. None of the charges~\eqref{Eqcharges} overlap the current, so there is no obvious zero-frequency Drude weight. Numerically, we find that the averaged Kubo correlators $C_{JJ}(t)+C_{JJ}(t+1)$ decays as $t^{-\beta}$, with an exponent $\beta \approx 1.7 >1$, indicating a finite d.c.~conductivity $\sigma(\omega=0)$, and thus a finite diffusion constant. The structure factor $C(x,t) = \langle q(x,t) q(0,0) \rangle$, with $q$ the local particle number appropriately coarse-grained over unit cells~\cite{SM}, displays an ever richer structure (Fig.~\ref{fig:current}), with some ballistic peak carrying vanishing weight due to $\sigma$-strings, and an asymmetric non-Gaussian diffusive peak near the origin. 

\par

	   	\emph{Discussion}.---In this work we introduced a new reversible cellular automaton based on the Fredkin update rule. We showed that the spectrum of this automaton contains two genera of stable quasiparticles, the $\beta$-strings and the $\sigma$ quasiparticles. The $\beta$-strings of all sizes have the same bare velocity, but are renormalized differently through their collisions with $\sigma$ quasiparticles. Thus this model features an infinite hierarchy of quasiparticles with distinct effective velocities above a typical thermal state. The motion of the $\sigma$ quasiparticles, meanwhile, is unaffected by the scattering, so it is not entirely clear if (and how) one can assign them to ``strings.'' As we discussed, the $\beta-\sigma$ scattering depends nontrivially on the spacing between adjacent $\sigma$ quasiparticles; while this dependence can be computed, we have not been able to factor it into contributions due to a hierarchy of $\sigma$-type strings. Thus the full Bethe ansatz solution of this model remains a task for future work. We remark that this model does not appear to fall under a current partial classification of integrable CAs~\cite{Gombor:2021ww,BalazsPrivate}.
		
		Although we were unable to fully solve the model, we could analytically establish the presence of an infinite hierarchy of conserved charges; physically, these charges represent the spacings between $\sigma$ quasiparticles, which are conserved. 
	Such asymptotic spacings are also conserved in the Rule 54 RCA~\cite{Gopalakrishnan:2018wj,Bu_a_2021} but do not seem to affect the hydrodynamics of the model. However, scattering of $\beta$ strings depends on spacings of $\sigma$-particles in a $\sigma$-string suggesting that they might play a role in determining the late time behavior of the FSA.
	\par
	Finally, we studied transport properties in this model. We found that the d.c. limit of transport is diffusive, but with an asymmetric and non-gaussian dynamic structure factor. Moreover, the model features persistent current oscillations, leading to a finite-frequency delta-function peak in the a.c. conductivity. 
	A comprehensive understanding of the transport behavior in this model should be amenable to generalized hydrodynamics (GHD)~\cite{Bastianello_2022}. However, this would require one to re-express the scattering data in a standard Bethe-ansatz form; this task is currently out of reach.  \par
	
\emph{Acknowledgements}.--- We thank B.~Pozsgay and B.~Ware for stimulating discussions.
This work was supported by the National Science Foundation under NSF Grant No. DMR-1653271 (S.G.), the US Department of Energy, Office of Science, Basic Energy Sciences, under Early Career Award No. DE-SC0019168 (R.V.), and the Alfred P. Sloan Foundation through a Sloan Research Fellowship (R.V.).

\bibliography{refs}

\bigskip
\onecolumngrid
\newpage


\section*{Supplementary material}

# Supplemental Material for "The Fredkin staircase: An integrable system with a finite-frequency Drude peak"

Hansveer Singh,[1] Romain Vasseur,[1] and Sarang Gopalakrishnan[2]

[1]*Department of Physics, University of Massachusetts Amherst, Massachusetts 01003, USA*
[2]*Department of Physics, Pennsylvania State University, University Park PA 16802, USA*
(Dated: May 17, 2022)

## 1. PROPERTIES OF QUASIPARTICLES

Owing to the integrable nature of the model, one could try to apply generalized hydrodynamics (GHD) [1] to study properties of the model. Such a program would require one to determine the density of quasiparticles of a given genus and species, the scattering shifts between said quasiparticles, and their bare velocities. In principle these desiderata should completely determine the dynamics of the model. As stated in the main text although we know the bare velocities, we were unable to obtain all of these quantities mostly due to the large number of species in the $\sigma$-string genus. In this section we present a few key ingredients we were able to extract numerically.

The first quantity we attempt to extract is the density of $\beta$-strings in a random state. As mentioned in the main text, the velocities of $\beta$-strings are renormalized when colliding with $\sigma$-strings. We found that one can use the following procedure to extract $\beta$-strings from a random state. First prepare the initial state (working in the occupation basis),

$$|00\cdots 00\rangle|\text{random state}\rangle|00\cdots 00\rangle|\sigma\text{-string with random spacings}\rangle|00\cdots 00\rangle, \qquad (1.1)$$

where the random state corresponds to a random configuration of 0s and 1s (i.e. a random bit string) and we denote the size of the region that the random bit string state occupies as $\ell$ and the size of the $\sigma$-string as $s$ and the spacing between the two regions or equivalently the size of the vacuum region between the two regions as $d$.

The time evolution of the state can be seen in the bottom left panel of Fig. 1. One can see that the random state and $\sigma$-string start to expand into the vacuum regions and after some time one sees that the $\sigma$ and $\beta$-strings from the random state separate from each other. Even though the $\beta$-strings and $\sigma$-strings are well separated from each other, the $\beta$-strings are not simply consecutive $\beta$-particles and so we would be unable to identify $\beta$-strings at this stage. However, at a later time the $\beta$-strings will collide with the randomly spaced size $s$ $\sigma$-string, their velocities will become renormalized when they travel inside the $\sigma$-string and if we make $s \gg \ell$ then we will be able to efficiently distinguish $\beta$-strings even though some $\beta$-strings such as $\beta$-particles will still be spatially close together after this filtration process.

Performing this procedure for many random initial states, we can obtain the density of $\beta$-strings. We present our results in the bottom right panel of Fig. 1. We can see that for string sizes, $s$, roughly up to $s = 30$ there seems to be power law behavior,$s^{-\gamma}$,with a $2 \lesssim \gamma \lesssim 3$. The behavior is difficult to ascertain for larger $\beta$-string sizes as this requires both more averaging and also larger system sizes for the random state so that there is a higher chance of observing such large $\beta$-strings. Indeed, in the bottom right panel of Fig. 1 one can see different qualitative behavior for in the distribution of $\beta$-strings at large enough $s$ for different system sizes.

The second quantity we we were able to obtain are scattering shifts between certain $\sigma$-strings and $\beta$-strings. We numerically determined from small string sizes the scattering shift a $\beta$-string of size $\ell$ receives when scattering off a $s$-spaced size $m$ type-A $\sigma$-string and were able to extrapolate a general expression. We found

$$\Delta x(\ell, s, m) = \begin{cases} m\ell & s \neq 0 \text{ and } \ell \leq 2s \\ \ell + 2(m-1)s & s \neq 0 \text{ and } \ell > 2s \\ \ell & s = 0. \end{cases} \qquad (1.2)$$

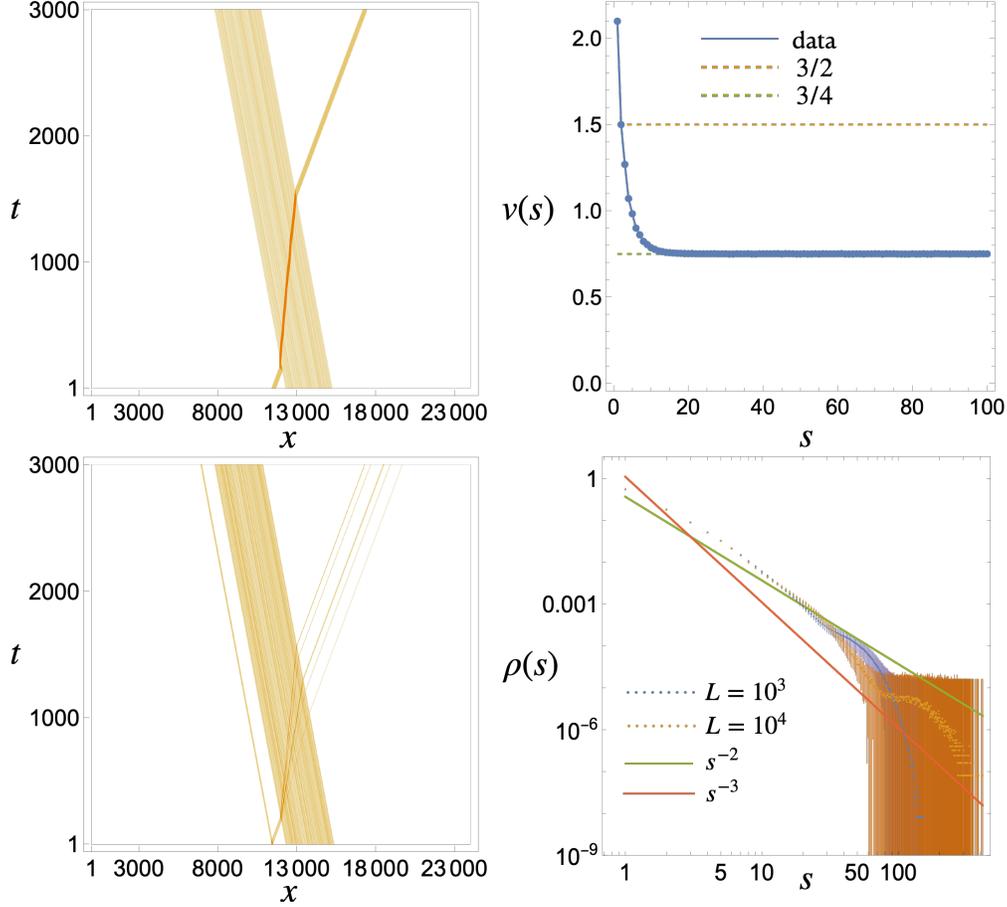

FIG. 1. **Vacuum expansion experiments.** *Top Left*: A sample vacuum expansion experiment of a $\beta$-string of size 40 colliding with a randomly spaced $\sigma$-string of size 200. From this run we extract the renormalized velocity of the $\beta$-string when it is inside the randomly spaced $\sigma$-string. *Top Right*: Renormalized velocity of $\beta$-strings when colliding with randomly spaced $\sigma$-strings which are much larger than the $\beta$-string size. One can see that the velocities are all lower than the bare velocity $v_\beta = 3$ and that saturation appears to occur for large $\beta$-strings to a value of $3/4$. *Bottom Left*: A sample vacuum expansion experiment where a random state is spaced far away from a large $\sigma$-string. The $\beta$-strings from the random state move to the right and eventually collide and are filtered by the $\sigma$-string. One can see that after leaving the $\sigma$-string, the $\beta$-strings are ordered from smallest to largest. From here we can extract the number of size-$s$ $\beta$-strings that were in the random state. *Bottom Right*: The density of $\beta$-strings extracted by the vacuum expansion experiment. Error bars cause the value to go below zero which is why they appear so large on the log-log scale. One can see that the density $\rho(s)$ appears to have a power law behavior with an exponent between 2 and 3.

Fig. 2 shows some collisions for between a $\beta$-string of size 6 and different $s$-spaced size $m$ type-A $\sigma$-strings. One can see that the formula predicts the scattering shift correctly but we note that this relation has not been proven analytically.

Finally we numerically extracted the velocity renormalization of bendy strings when encountering type A $\sigma$-strings which have random spacings. This was done in a similar fashion to extracting the $\beta$-string distributions by changing the initial state to

$$|00\cdots 00\rangle|\text{size } s \text{ } \beta\text{-string}\rangle|00\cdots 00\rangle|\sigma\text{-string with random spacings}\rangle|00\cdots 00\rangle. \tag{1.3}$$

Knowing the bare velocity of a $\beta$-string is 3 and the bare velocity of a $\sigma$-string is $3/2$ one can determine the position and time when the rightmost $\beta$-particle in the $\beta$-string will first meet the first $\sigma$-particle in the $\sigma$-

2

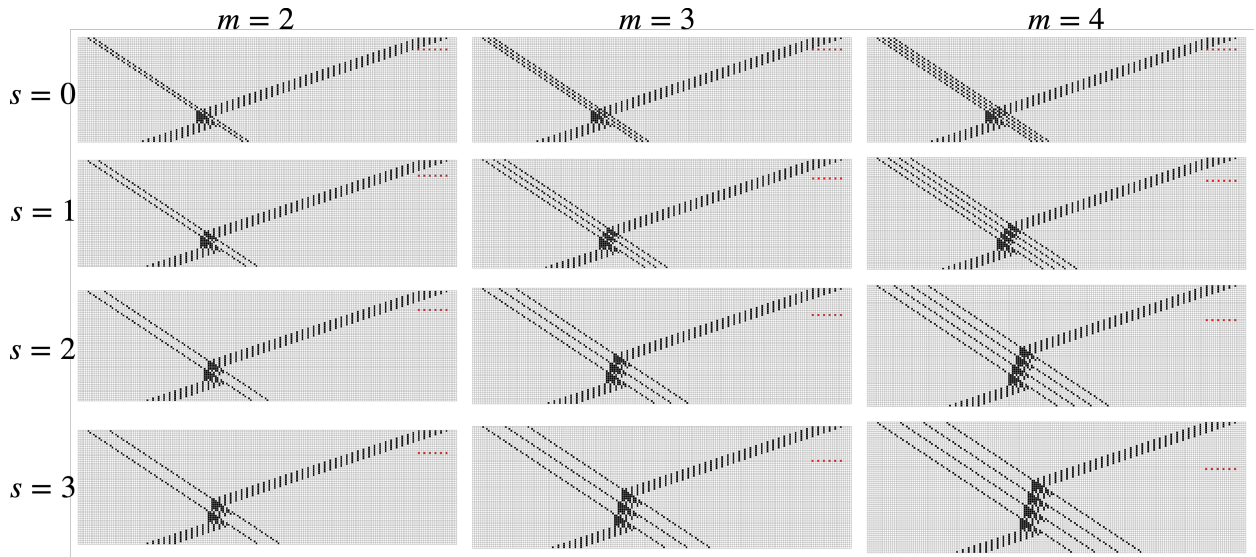

FIG. 2. **Scattering events.** A $\beta$-string of size 6 collides with type A $\sigma$-strings which are $s$-spaced and are of size $m$. Time runs vertically upward and space runs horizontally to the right. The red squares represents where the $\beta$ string would have been at the same time $t$. One can see that all the scattering shifts match with Eq. 1.2.

string. Similarly one can view the experiment backwards in time and determine the position and time when the rightmost $\sigma$-particle of the $\sigma$-string meets with the first $\beta$-particle of the $\beta$-string. This will correspond to the position and time at which the $\beta$-string exits. Calculating these quantities gives the duration for which the $\beta$-string was traveling in the $\sigma$-string and the distance it travelled—hence one can calculate the renormalized velocity of the $\beta$-string. The behavior of the renormalized velocity is shown in the top right panel of Fig. 1. While we currently are unable to explain the behavior of the renormalized velocity, we point out that all strings appear to be renormalized to a velocity lower than 3 and larger strings appear to have a renormalized velocity which saturates to a value of 3/4.

## 2. NONINTEGRABLE FREDKIN RCA

Integrability, and hence having quasiparticles with infinite lifetimes, is a fine-tuned phenomena and can usually be broken by making small deformations to the dynamics. Here we demonstrate that changing the gate pattern outlined in the main text is enough to break the integrability of the FSA. The gate pattern for a single Floquet step we used is shown in the top left panel of Fig. 3. In formulas the Floquet operator, $\mathcal{U}$, is now given by $\mathcal{U} = V_4 V_2 V_3 V_1$, where

$$V_i = \prod_{j \equiv i \bmod 4} U_{j,j+1,j+2,j+3} \tag{2.1}$$

and the expression for $U_{j,j+1,j+2,j+3}$ is the same as in the main text. We show a typical trajectory in the bottom left panel of Fig. 3 where one can identify backscattering events of quasiparticles—an example is shown in the bottom right panel of Fig. 3. Such backscattering events demonstrate that the model is nonintegrable since this implies that three body scattering events due not factorize. Furthermore, if the model is nonintegrable then one expects based on the results of Ref. [2] that the model should show subdiffusive transport with an exponent $z \simeq 8/3$. We show that the structure factor, $C(x,t) = \langle q_x(t) q_0 \rangle$, indeed shows a nice scaling collapse to $z = 8/3$ using the usual scaling form $C(x,t) = t^{1/z} f(x t^{-1/z})$. Thus this gives another example in confirming the universality of the exponent $z = 8/3$ for nonintegrable models evolving under the Fredkin constraint.

3

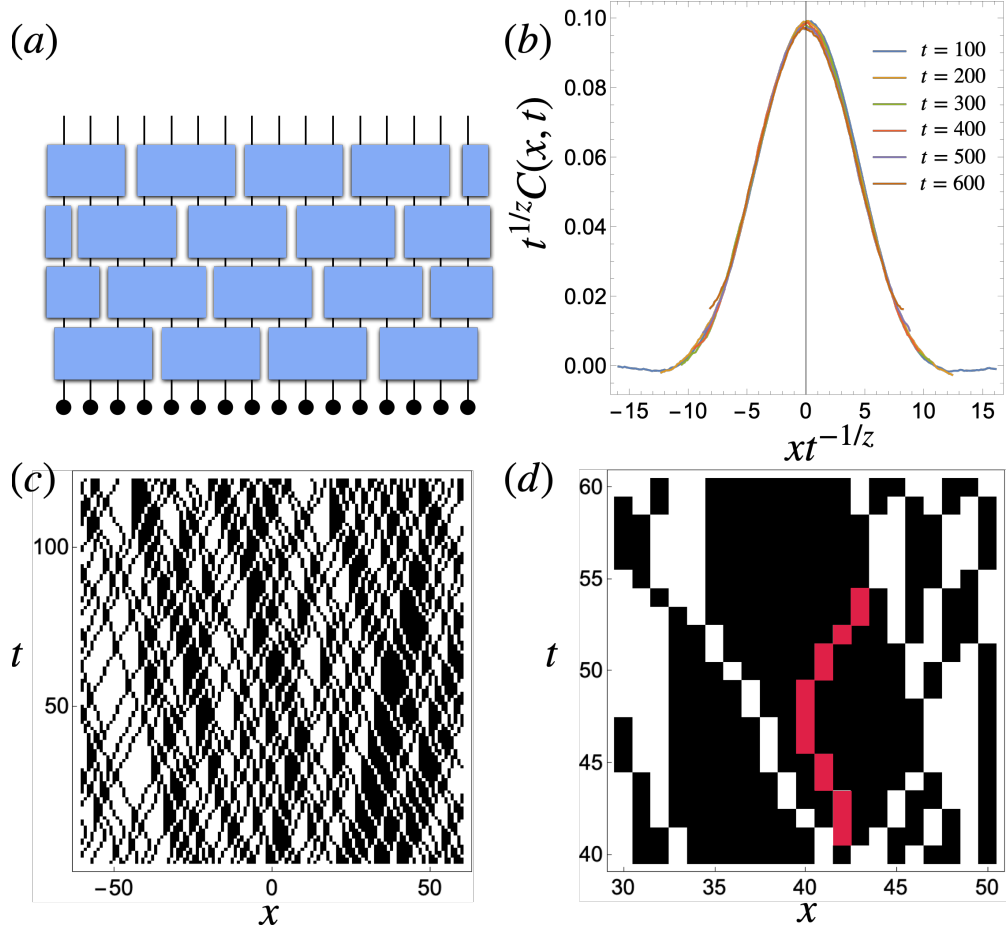

FIG. 3. **Nonintegrable Fredkin CA.** (a) Picture of the gate pattern for the nonintegrable Fredkin automaton. As opposed to the gate pattern in the main text, we have four layers rather than three. (b) Structure factor of the nonintegrable Fredkin automaton which shows a good scaling collapse with $z = 8/3$. A moving average over 20 sites was performed to reduce noise in the data. (c) A typical evolution of the automaton under the dynamics of the gates in Eq. 2.1.(d) A close up of a backscattering event which occurs in panel (c) demonstrating the model is not integrable. The quasiparticle which backscatters is shown in red.

## 3. NUMERICAL METHODS

Since the model is a reversible cellular automaton (RCA), one only needs to evolve product states in the occupation basis, i.e. the local Hilbert space at site $x$ being spanned by the basis $|\circ\rangle_x$ and $|\bullet\rangle_x$ for which the local charge is $n_x$ which has eigenvalues 0 and 1. Equivalently, in the spin basis, i.e. the local Hilbert space at site $x$ being spanned by $|\downarrow\rangle$ and $|\uparrow\rangle$ for which the local charge is $Z_x$, the usual Pauli-$z$ operator on site $x$. The two bases are related to each other via the relations: $|\bullet\rangle = |\uparrow\rangle$, $|\circ\rangle = |\downarrow\rangle$ and $n_x = \frac{1+Z_x}{2}$. Practically speaking, for a system of size $L$ one creates an array with $L$ entries given by either 0s or 1s (occupation basis) or with entries given by 1s and -1s (spin basis) to specify the initial state and then one updates the entries of this array according to the FSA rules with the corresponding gate geometry given in the main text. Since RCA send product states to product states in a particular basis, one only needs to use the above array to access all information about the time evolution of a given product state.

4

### 3.1. Computing Correlation Functions

Our main objective is to compute unequal time connected correlation functions at infinite temperature and zero chemical potential, i.e $\langle O(t)O(0)\rangle - \langle O(t)\rangle\langle O'(0)\rangle$, where $O$ and $O'$ are diagonal in the spin basis and $\langle \cdot \rangle \equiv \frac{1}{2^L}\text{tr}(\cdot)$. Expanding out the trace in the spin basis one has,

$$\langle O(t)O(0)\rangle - \langle O(t)\rangle\langle O'(0)\rangle = \frac{1}{2^L}\left(\sum_\sigma O_\sigma(t)O'_\sigma(0)\right) - \frac{1}{4^L}\left(\sum_{\sigma,\tau} O_\sigma(t)O'_\tau(0)\right) \quad (3.1)$$

where $\sigma$ and $\tau$ denote spin configurations and $O_\sigma(t) \equiv \langle\sigma|O(t)|\sigma\rangle = \langle\sigma(t)|O(0)|\sigma(t)\rangle$.

The two correlation functions we computed are the current-current correlation function, $C_{JJ}(t) = \frac{1}{L}\langle J(t)J(0)\rangle$, and the structure factor, $C_{ZZ}(t) = \langle Z_x(t)Z_0(0)\rangle$, where $J = \sum_x j_x$ is the total current and $Z_0$ is the Pauli-$z$ matrix in the middle of the chain. As we will show in the next section, $J$ is traceless so we do not have to compute the disconnected piece for either of these correlation functions. To compute these quantities we have to sample states, $|\sigma\rangle$, then evolve these states to time $t$ with periodic boundary conditions (PBC) and then evaluate $Z_{0,\sigma}(t)$ or $J_\sigma(t)$. If we uniformly sample $M$ states, then the error in the correlation functions should go down with $\sqrt{M}$ by the central limit theorem. We checked convergence with sample size for the both of these correlations but we only show the convergence check for the current-current correlator in the top left panel of Fig. 4.

Lastly, we remark that computing $C_{ZZ}(t)$ is equivalent to computing $\langle Z_x(t)\rangle_{Z_0=1}$ where the $\langle \cdot \rangle_{Z_0=1}$ indicates only averaging over states, $\sigma$, with $Z_{0,\sigma} = 1$. To see this, we compute $C_{nn}(t) \equiv \langle n_x(t)n_0(0)\rangle - \langle n_x(t)\rangle\langle n_0(0)\rangle$ in the occupation basis. We have,

$$C_{nn}(t) = \frac{1}{2^L}\left(\sum_m n_{x,m}(t)n_{0,m}(0)\right) - \langle n_x(t)\rangle\langle n_0(t)\rangle \quad (3.2)$$

where $m$ denotes occupation number configurations and $n_{x,m}(t) \equiv \langle m(t)|n_{x,m}(0)|m(t)\rangle$. Since $n_x$ only has eigenvalues 0 or 1 then the above sum only runs over configurations such that $n_{0,m}(0)=1$ which is equivalent to $Z_{0,m}(0) = 1$. Translating back to the spin basis we have,

$$C_{nn}(t) = \frac{1}{2^L}\left(\sum_{\sigma|Z_{0,\sigma}(0)=1} \frac{1+Z_{x,m}(t)}{2}\right) - \langle \frac{1+Z_x(t)}{2}\rangle\langle\frac{1+Z_0(0)}{2}\rangle$$

$$= \frac{1}{4}\sum_{\sigma|Z_{0,\sigma}(0)=1} Z_{x,m}(t).$$

However, we also have that $C_{nn}(t) = \frac{1}{4}C_{ZZ}(t)$ by using the relations specified earlier between the spin and occupation basis. Thus,

$$C_{ZZ}(t) = \sum_{\sigma|Z_{0,\sigma}(0)=1} Z_{x,m}(t). \quad (3.3)$$

This simplification doesn't occur for $C_{JJ}(t)$ since the current has a support larger than one (shown in the next section). Thus we need to keep track of the initial value for the current when computing the correlation function. In the remainder of the supplementary material we will use $q_x(t)$ instead of $Z_x(t)$.

### 3.2. TEBD Simulations

A pitfall to the above method of simulation is that one has to sample many states (typically of order $10^8$) in order to obtain clean data at long times. One can overcome this by switching to evolving the current operator, $J(0)$, or the charge operator at site 0, $q_0(0)$, using time-evolving block decimation (TEBD) with matrix product operators (MPO) [3] and then one can compute unequal time correlation functions with the

5

exact infinite temperature density matrix by performing a simple contraction between the MPO at time $t$ and zero, i.e. one can view $\frac{1}{2^L}\text{tr}(J(t)J(0))$ as an inner product by vectorizing $J(t)$ into the state $|J(t)\rangle\rangle$ and then compute $\langle\langle J(t)|J(0)\rangle\rangle$ which is simply an overlap between two states which can be done efficiently with TEBD, where $\langle\langle A|B\rangle\rangle \equiv \frac{1}{2^L}\text{tr}(A^\dagger B)$.

For this method to be useful, one has to have a low bond dimension MPO representation of $q_0(0)$ and $J(0)$. While the former is straightforwardly seen to be a bond dimension one MPO, the bond dimension of the total current is not as simple to ascertain since the current—as we show in the next section—has a rather unwieldy expression. However, we will show that one only needs access to an efficient MPO representation of the coarse grained unit cell current, $j_x^{\text{unit}}(0)$, which we define momentarily, and one needs to compute expectation values, $\langle j_0^{\text{unit}}(t) j_x^{\text{unit}}(0)\rangle$, when $j_x^{\text{unit}}(0)$ has overlapping support with $j_0^{\text{unit}}(t)$. In our work we perform TEBD simulations with open boundary conditions (OBC), so we have to compute

$$C_{JJ}(t) = \frac{1}{L} \sum_{x,y=-L/2}^{L/2} \langle j_x(t) j_y(0)\rangle. \tag{3.4}$$

The local currents have finite support and cannot grow faster than the circuit light cone set by the geometry show in the main text. As such we only need to perform the this expectation value over the region of space where the tensors of the MPO representation of $j_x(t)$ are not equal to 1. Denoting this region by $A$ and its size by $\ell$, we have

$$C_{JJ}(t) = \frac{1}{L} \sum_{x,y \in A} \langle j_x(t) j_y(0)\rangle. \tag{3.5}$$

In this region the system effectively behaves as if it is in the thermodynamic limit despite the use of OBC and hence we can make use of the three site translation invariance of the FSA. To make use of this invariance we need to introduce the unit cell current, $j_x^{\text{unit}}(t)$, whose expression is

$$j_x^{\text{unit}}(t) \equiv \frac{1}{3}(j_x(t) + j_{x+1}(t) + j_{x+2}(t)), \tag{3.6}$$

and here we choose the convention that $x \equiv 0 \bmod 3$. The current correlator becomes,

$$C_{JJ}(t) = \frac{9}{L} \sum_{x,y \in A | x,y \equiv 0 \bmod 3} \langle j_x^{\text{unit}}(t) j_y^{\text{unit}}(0)\rangle. \tag{3.7}$$

Making use of the the three-site translational invariance, we get

$$C_{JJ}(t) = 3 \sum_{y \in A | y \equiv 0 \bmod 3} \langle j_0^{\text{unit}}(t) j_y^{\text{unit}}(0)\rangle. \tag{3.8}$$

Thus one only needs to calculate the overlap of $j_0^{\text{unit}}(t)$ with other local currents that are within the light cone of $j_0^{\text{unit}}(t)$. This light cone can be determined easily as one simply searches for the farthest right and left tensor that is equal to the identity. Finally, using Eq. 4.3–4.5 we can numerically write down $j_0^{\text{unit}}(0)$ as a matrix since its support is only over 10 sites so it is a $1024 \times 1024$ matrix. One then turns this matrix into an MPO representation by repeatedly applying successive SVDs [3] and one finds an MPO representation with maximum bond dimension 11. We note that one could possibly also obtain an exact analytical expression of the MPO tensors by using finite weighted automaton methods[4].

We show the comparison between TEBD and classical sampling in the bottom left figure of Fig. 4. One can see at early times TEBD and the classical simulations agree but fail to agree with each other at later times. However, we do see that if one increases the maximum bond dimension then TEBD simulations start to converge to classical simulations. This is to be expected as one typically encounters an entanglement barrier due to unitary evolution. Obviously, the only limitation to the accuracy of the classical simulations is the amount of states one samples and as we have shown we have sampled enough states to achieve good convergence.

## 4. CURRENT CALCULATION

The total current is given by $J(t) = \sum_x j_x(t)$ and $j_x(t)$ is the local current density. To calculate $j_x(0)$, one makes use of the continuity equation,

$$q_x(t+1) - q_x(t) + j_{x+1}(t) - j_x(t) = 0. \tag{4.1}$$

One can use the above continuity equation to construct a telescoping sum, so that the current terms evaluated at $t=0$ which survive have orthogonal support. More concretely we have,

$$\sum_{x=x_0}^{x_1} q_x(1) - q_x(0) = \sum_{x=x_0}^{x_1} j_{x+1}(0) - j_x(0) = j_{x_1+1}(0) - j_{x_0}(0). \tag{4.2}$$

As long as $|x_1 - x_0| \gg \max(\operatorname{supp}(j_{x_0}(0)), \operatorname{supp}(j_{x_1+1}(0)))$—where $\operatorname{supp}(j_x(0))$ refers to the size of the support of $j_x(t)$— then $j_{x_1+1}(0)$ and $j_{x_0}(0)$ can be distinguished since their support do not overlap.

The calculation is rather cumbersome since one has to compute $q_x(1)$ and this is a tedious task since the unitary gates have large support and we also must apply a full Floquet step, thus we performed the computation in Mathematica. We found that the form of the current density, $j_x(0)$, depends on whether $x \equiv 0 \bmod 3, 1 \bmod 3$, or $2 \bmod 3$. Their lengthy expressions read (omitting the time argument in the charges for readability),

$$j_x(0) = \frac{1}{4}(q_{x-1} - q_x)(-q_{x+1} + q_{x-2}(q_{x+1} + 1) + 3), \quad x \equiv 0 \bmod 3 \tag{4.3}$$

$$\begin{aligned}
j_x(0) = \frac{1}{16}(& q_{x+2}q_{x+3} + q_{x+1}q_x + 3q_{x+2}q_x + q_{x+1}q_{x+3}q_x \\
& - q_{x+2}q_{x+3}q_x - 12q_x - q_{x+1} + q_{x+2} - q_{x+1}q_{x+3} \\
& - q_{x-1}(4q_{x-3}(q_x + 1) + 3q_xq_{x+2} + q_{x+2} - q_xq_{x+2}q_{x+3} \\
& + q_{x+2}q_{x+3} + (q_x - 1)q_{x+1}(q_{x+3} + 1) - 4) + q_{x-2}(q_{x+1}q_x \\
& - q_{x+2}q_x + q_{x+1}q_{x+3}q_x - q_{x+2}q_{x+3}q_x - 4q_x + 4q_{x-3} \\
& \times (q_x + 1) - q_{x+1} - 3q_{x+2} - q_{x+1}q_{x+3} + q_{x+2}q_{x+3} + q_{x-1} \\
& \times (-((q_x - 1)q_{x+1}(q_{x+3} + 1)) + q_{x+2} \\
& \times (-q_{x+3} + q_x(q_{x+3} + 1) + 3) + 4) + 8)), \quad x \equiv 1 \bmod 3
\end{aligned} \tag{4.4}$$

$$\begin{aligned}
j_x(0) = \frac{1}{64}(&8q_xq_{x-1} - 4q_xq_{x+1}q_{x-1} - 4q_{x+1}q_{x-1} + 4q_xq_{x+2}q_{x-1} - 16q_{x+1}q_{x+2}q_{x-1} - 4q_{x+2}q_{x-1} \\
&+ q_xq_{x+3}q_{x-1} - q_xq_{x+1}q_{x+3}q_{x-1} - q_{x+1}q_{x+3}q_{x-1} - q_xq_{x+2}q_{x+3}q_{x-1} + q_xq_{x+1}q_{x+2}q_{x+3}q_{x-1} \\
&+ q_{x+1}q_{x+2}q_{x+3}q_{x-1} - q_{x+2}q_{x+3}q_{x-1} + q_{x+3}q_{x-1} + 3q_xq_{x+4}q_{x-1} - 3q_xq_{x+1}q_{x+4}q_{x-1} + q_{x+1}q_{x+4}q_{x-1} \\
&+ q_xq_{x+2}q_{x+4}q_{x-1} - q_xq_{x+1}q_{x+2}q_{x+4}q_{x-1} + 3q_{x+1}q_{x+2}q_{x+4}q_{x-1} - 3q_{x+2}q_{x+4}q_{x-1} - q_{x+4}q_{x-1} \\
&+ q_xq_{x+3}q_{x+5}q_{x-1} - q_xq_{x+1}q_{x+3}q_{x+5}q_{x-1} - q_{x+1}q_{x+3}q_{x+5}q_{x-1} - q_xq_{x+2}q_{x+3}q_{x+5}q_{x-1} \\
&+ q_xq_{x+1}q_{x+2}q_{x+3}q_{x+5}q_{x-1} + q_{x+1}q_{x+2}q_{x+3}q_{x+5}q_{x-1} - q_{x+2}q_{x+3}q_{x+5}q_{x-1} + q_{x+3}q_{x+5}q_{x-1} \\
&- q_xq_{x+4}q_{x+5}q_{x-1} + q_xq_{x+1}q_{x+4}q_{x+5}q_{x-1} + q_{x+1}q_{x+4}q_{x+5}q_{x-1} + q_xq_{x+2}q_{x+4}q_{x+5}q_{x-1} \\
&- q_xq_{x+1}q_{x+2}q_{x+4}q_{x+5}q_{x-1} - q_{x+1}q_{x+2}q_{x+4}q_{x+5}q_{x-1} + q_{x+2}q_{x+4}q_{x+5}q_{x-1} - q_{x+4}q_{x+5}q_{x-1} \\
&+ 16q_{x-1} - 8q_x + 4q_xq_{x+1} - 44q_{x+1} - 4q_xq_{x+2} + 16q_{x+1}q_{x+2} + 4q_{x+2} \\
&- 4q_{x-2}(4q_{x-4}(q_{x-1}+1) + 3q_{x-1}q_{x+1} + q_{x+1} - q_{x-1}q_{x+1}q_{x+2} + q_{x+1}q_{x+2} + (q_{x-1}-1)q_x(q_{x+2}+1) - 4) \\
&- q_xq_{x+3} + q_xq_{x+1}q_{x+3} + q_{x+1}q_{x+3} + q_xq_{x+2}q_{x+3} - q_xq_{x+1}q_{x+2}q_{x+3} - q_{x+1}q_{x+2}q_{x+3} + q_{x+2}q_{x+3} \\
&- q_{x+3} - 3q_xq_{x+4} + 3q_xq_{x+1}q_{x+4} - q_{x+1}q_{x+4} - q_xq_{x+2}q_{x+4} + q_xq_{x+1}q_{x+2}q_{x+4} - 3q_{x+1}q_{x+2}q_{x+4} \\
&+ 3q_{x+2}q_{x+4} + q_{x+4} - q_xq_{x+3}q_{x+5} + q_xq_{x+1}q_{x+3}q_{x+5} + q_{x+1}q_{x+3}q_{x+5} + q_xq_{x+2}q_{x+3}q_{x+5} \\
&- q_xq_{x+1}q_{x+2}q_{x+3}q_{x+5} - q_{x+1}q_{x+2}q_{x+3}q_{x+5} + q_{x+2}q_{x+3}q_{x+5} - q_{x+3}q_{x+5} + q_xq_{x+4}q_{x+5} \\
&- q_xq_{x+1}q_{x+4}q_{x+5} - q_{x+1}q_{x+4}q_{x+5} - q_xq_{x+2}q_{x+4}q_{x+5} + q_xq_{x+1}q_{x+2}q_{x+4}q_{x+5} + q_{x+1}q_{x+2}q_{x+4}q_{x+5} \\
&- q_{x+2}q_{x+4}q_{x+5} + q_{x+4}q_{x+5} + q_{x-3}(16q_{x-4}(q_{x-1}+1) + 8q_{x-1}q_x - 8q_x - 4q_{x-1}q_{x+1} - 4q_{x-1}q_xq_{x+1} \\
&+ 4q_xq_{x+1} - 12q_{x+1} + 12q_{x-1}q_{x+2} + 4q_{x-1}q_xq_{x+2} - 4q_xq_{x+2} - 12q_{x+2} \\
&- 4q_{x-2}((q_{x-1}-1)q_x(q_{x+2}+1) - q_{x+1}(-q_{x+2}+q_{x-1}(q_{x+2}+1)+3) - 4) + q_{x-1}q_{x+3} + q_{x-1}q_xq_{x+3} \\
&- q_xq_{x+3} - q_{x-1}q_{x+1}q_{x+3} - q_{x-1}q_xq_{x+1}q_{x+3} + q_xq_{x+1}q_{x+3} + q_{x+1}q_{x+3} - q_{x-1}q_{x+2}q_{x+3} \\
&- q_{x-1}q_xq_{x+2}q_{x+3} + q_xq_{x+2}q_{x+3} + q_{x-1}q_{x+1}q_{x+2}q_{x+3} + q_{x-1}q_xq_{x+1}q_{x+2}q_{x+3} - q_xq_{x+1}q_{x+2}q_{x+3} \\
&- q_{x+1}q_{x+2}q_{x+3} + q_{x+2}q_{x+3} - q_{x+3} - q_{x-1}q_{x+4} + 3q_{x-1}q_xq_{x+4} - 3q_xq_{x+4} + q_{x-1}q_{x+1}q_{x+4} \\
&- 3q_{x-1}q_xq_{x+1}q_{x+4} + 3q_xq_{x+1}q_{x+4} - q_{x+1}q_{x+4} - 3q_{x-1}q_{x+2}q_{x+4} + q_{x-1}q_xq_{x+2}q_{x+4} - q_xq_{x+2}q_{x+4} \\
&+ 3q_{x-1}q_{x+1}q_{x+2}q_{x+4} - q_{x-1}q_xq_{x+1}q_{x+2}q_{x+4} + q_xq_{x+1}q_{x+2}q_{x+4} - 3q_{x+1}q_{x+2}q_{x+4} + 3q_{x+2}q_{x+4} + q_{x+4} \\
&+ q_{x-1}q_{x+3}q_{x+5} + q_{x-1}q_xq_{x+3}q_{x+5} - q_xq_{x+3}q_{x+5} - q_{x-1}q_{x+1}q_{x+3}q_{x+5} - q_{x-1}q_xq_{x+1}q_{x+3}q_{x+5} \\
&+ q_xq_{x+1}q_{x+3}q_{x+5} + q_{x+1}q_{x+3}q_{x+5} - q_{x-1}q_{x+2}q_{x+3}q_{x+5} - q_{x-1}q_xq_{x+2}q_{x+3}q_{x+5} + q_xq_{x+2}q_{x+3}q_{x+5} \\
&+ q_{x-1}q_{x+1}q_{x+2}q_{x+3}q_{x+5} + q_{x-1}q_xq_{x+1}q_{x+2}q_{x+3}q_{x+5} - q_xq_{x+1}q_{x+2}q_{x+3}q_{x+5} - q_{x+1}q_{x+2}q_{x+3}q_{x+5} \\
&+ q_{x+2}q_{x+3}q_{x+5} - q_{x+3}q_{x+5} - q_{x-1}q_{x+4}q_{x+5} - q_{x-1}q_xq_{x+4}q_{x+5} + q_xq_{x+4}q_{x+5} + q_{x-1}q_{x+1}q_{x+4}q_{x+5} \\
&+ q_{x-1}q_xq_{x+1}q_{x+4}q_{x+5} - q_xq_{x+1}q_{x+4}q_{x+5} - q_{x+1}q_{x+4}q_{x+5} + q_{x-1}q_{x+2}q_{x+4}q_{x+5} + q_{x-1}q_xq_{x+2}q_{x+4}q_{x+5} \\
&- q_xq_{x+2}q_{x+4}q_{x+5} - q_{x-1}q_{x+1}q_{x+2}q_{x+4}q_{x+5} - q_{x-1}q_xq_{x+1}q_{x+2}q_{x+4}q_{x+5} + q_xq_{x+1}q_{x+2}q_{x+4}q_{x+5} \\
&+ q_{x+1}q_{x+2}q_{x+4}q_{x+5} - q_{x+2}q_{x+4}q_{x+5} + q_{x+4}q_{x+5} + 16)), \quad x \equiv 2 \bmod 3. \end{aligned} \qquad (4.5)$$

We checked (not shown) that the continuity equation is satisfied for every point for individual realizations. Another non-trivial relation which we can check is that the current-current correlator, $C_{JJ}(t)$, is related to the second (finite difference) derivative of the charge-charge correlation functions, i.e.

$$\frac{d^2\overline{x^2}}{dt^2} \equiv \overline{x^2}(t-1) - 2\overline{x^2}(t) + \overline{x^2}(t+1) = 2C_{JJ}(t). \qquad (4.6)$$

where $\overline{f}(t) \equiv \sum_x f(x)C(x,t)$ and $C(x,t)$ is a coarse-grained charge-charge correlator. Before proving this relation we need to define the correct coarse-grained charge-charge correlator, $C(x,t)$. Its expression is given below,

$$C_{\text{unit}}(x,t) = \frac{1}{9}\langle(q_x(t) + q_{x+1}(t) + q_{x+2}(t))(q_0(0) + q_1(0) + q_2(0))\rangle \qquad (4.7)$$

$$C(x,t) = \frac{1}{3}(C_{\text{unit}}(x,t) + C_{\text{unit}}(x+1,t) + C_{\text{unit}}(x+2,t)). \qquad (4.8)$$

8

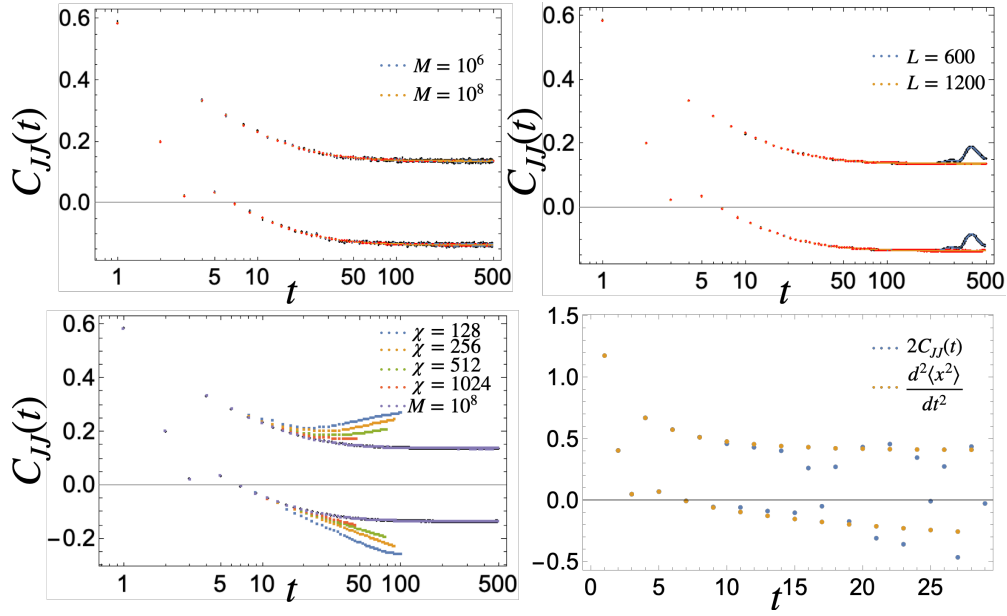

FIG. 4. **Transport and Kubo correlator.** *Top Left*: The current-current correlator, $C_{JJ}(t)$, computed for $L = 1200$ when averaged over $M$ initial states. One can see that results are converged when one increases the sample size by a factor of 100. *Top Right*: $C_{JJ}(t)$ computed for $L = 600$ and $L = 1200$. One can see that the two agree until $t \approx 200$, indicating the onset of finite size behavior for $L = 600$. *Bottom Left*: Comparison of computing $C_{JJ}(t)$ with TEBD simulations vs classical sampling. One can see that the two methods agree up to $t \approx 20$ and then show significant deviations. These deviations tend to grow smaller when one increases the maximum bond dimension, $\chi$, in the TEBD simulations suggesting that TEBD will converge to the classically simulated result. *Bottom Right*: We numerically check Eq. 4.6 using TEBD for short times to illustrate that we indeed have the correct current. TEBD has the advantage of having less noise occur so taking finite difference derivatives leads to less errors. One can see that the two observables agree very well up to $t \approx 7$ at which appoint one starts to see deviations. These deviations are most likely due to truncation errors in TEBD since one needs high accuracy to perform finite difference derivatives correctly.

We note that the above normalization factors lead to the sum rule, $\sum_x C(x,t) = 1$.

Let us go through the proof:

$$\overline{x^2}(t-1) - 2\overline{x^2}(t) + \overline{x^2}(t+1)$$
$$= \sum_x x^2 (C(x, t-1) - 2C(x, t) + C(x, t+1))$$
$$= -\frac{1}{9} \sum_x x^2 \langle (j_{x+3}^{\text{unit}}(t) - j_x^{\text{unit}}(t))(j_3^{\text{unit}}(0) - j_0^{\text{unit}}(0)) \rangle,$$

where $j_x^{\text{unit}}(t) = \frac{1}{3}(j_x(t) + j_{x+1}(t) + j_{x+2}(t))$. At this point it is worth remarking that the reason one gets $j_{x+3}^{\text{unit}}(t)$ rather than $j_{x+1}^{\text{unit}}(t)$ is due to our definition of $C(x,t)$. This is needed because the Floquet time evolution operator only commutes with translation by three sites instead of a single site. Moving forward we have,

$$-\frac{1}{9} \sum_x x^2 \langle (j_{x+3}^{\text{unit}}(t) - j_x^{\text{unit}}(t))(j_3^{\text{unit}}(0) - j_0^{\text{unit}}(0)) \rangle$$
$$= -\frac{1}{9} \sum_x (2x^2 - (x+3)^2 - (x-3)^2) \langle j_x^{\text{unit}}(t) j_0^{\text{unit}}(0) \rangle + \mathcal{O}(e^{-\ell})$$
$$= 2C_{JJ}(t),$$

9

where $\ell$ denotes the size of the region we are summing over. For classical simulations $\ell = L$, the system size, since we simulate with PBC and for TEBD $\ell$ is the size of the light cone since we can only use translation invariance in this region since we use OBC. We lastly remark that while Eq. 4.8 was used to ensure that we could make use of translational invariance, Eq. 4.7 is used so that we obtain the unit cell current rather than the current of a single site, $j_x(t)$, since $\sum_x \langle j_x(t) j_0(0) \rangle$ is not related to $C_{JJ}(t)$. We show how our TEBD results match with the result in the bottom right figure of Fig. 4. At very early times we see good agreement but at later times we see noticeable deviation which we attribute to both bond dimension truncation since performing a finite difference second derivative requires much more accuracy. Needless to say, seeing that both the continuity equation and that Eq. 4.6 are satisfied indeed show that we are dealing with the correct current.



\end{document}